\documentclass[twocolumn,amsmath,amssymb,superscriptaddress,prl,showpacs]{revtex4}

\usepackage{graphicx}
\usepackage{amsmath,amssymb,amsthm}

\renewcommand{\vec}[1]{\boldsymbol{#1}}
\newcommand{\beq}{\begin{equation}}
\newcommand{\eeq}{\end{equation}}
\newcommand{\beqa}{\begin{eqnarray}}
\newcommand{\eeqa}{\end{eqnarray}}
\newcommand{\e}{\mathrm{e}}
\newcommand{\w}{\omega}

\newcommand{\ket}[1]{\left| #1 \right\rangle}

\newcommand{\ketbra}[2]{\left|#1\right\rangle\hskip-1mm\left\langle #2\right|}

\newcommand{\kv}{ {\bf k} }

\begin{document}

\title{Model of the Optical Emission of a Driven Semiconductor Quantum Dot: Phonon-Enhanced Coherent Scattering and Off-Resonant Sideband Narrowing} 
\author{Dara P. S. McCutcheon} \email{daramcc@df.uba.ar}
\affiliation{\mbox{Departamento de F\'isica, FCEyN, UBA and IFIBA, Conicet, Pabell\'on 1,~Ciudad~Universitaria, 1428 Buenos Aires, Argentina}}
\affiliation{Blackett Laboratory, Imperial College London, London SW7 2AZ, United Kingdom}
\author{Ahsan Nazir}\email{a.nazir@imperial.ac.uk}
\affiliation{Blackett Laboratory, Imperial College London, London SW7 2AZ, United Kingdom}

\date{\today}

\begin{abstract}

We study the crucial role played by 
the solid-state environment in
determining the photon emission characteristics of a driven quantum
dot. For resonant driving, we predict a 
phonon-enhancement of 
the coherently emitted radiation field 
with increasing driving strength,
in stark contrast to the conventional expectation
of a rapidly decreasing fraction of coherent emission 
with stronger driving. This surprising 
behaviour results from thermalisation of the dot with respect to the phonon bath, and leads to a nonstandard regime of resonance fluorescence in which significant coherent scattering and the Mollow triplet 
coexist. 
Off-resonance, we show that despite the
phonon influence, narrowing of dot 
spectral sideband widths can 
occur in certain regimes, consistent with an 
experimental trend.

\end{abstract}


\maketitle

As described by Mollow, the spectrum 
of light scattered from a resonantly 
driven two-level 
system (TLS) depends crucially on the relative size of the laser driving strength to the TLS 
radiative decay rate~\cite{mollow69}. For weak driving, 
the light is predominately coherently (or elastically) scattered, resulting in a single (delta function) peak in the emission spectrum at the laser frequency. At larger driving strengths, however, coherent scattering is strongly suppressed, and the emission 
becomes dominated by 
incoherent (inelastic) scattering from the TLS-laser dressed states~\cite{carmichael}.~This results 
in a triple-peak structure in the spectrum, known as the Mollow triplet.

While these fundamental predictions have 
long been confirmed in the traditional quantum optical setting of driven atoms~\cite{schuda74}, interest has turned more recently to their observation in solid-state TLSs (artificial atoms) such as 
semiconductor quantum dots (QDs)~\cite{xu07,muller07,ates09,flagg09_short,vamivakas09,ulrich11_short,ulhaq12}, 
single molecules~\cite{wrigge08}, and superconducting circuits~\cite{astafiev10}. %In fact, 
In the particular case of QDs, 
many of the archetypal features of  
atomic quantum optics have now been demonstrated, such as resonance fluorescence~\cite{xu07,muller07,ates09,flagg09_short,vamivakas09,ulrich11_short,ulhaq12}, 
coherent population oscillations~\cite{zrenner02,flagg09_short,ramsay10,ramsay10_2}, photon anti-bunching~\cite{michler00,santori01}, and two-photon 
interference~\cite{santori02,flagg10,patel10}. Aside from being of fundamental 
interest, these observations also 
pave the way towards using QDs as efficient single photon sources~\cite{nguyen11,matthiesen12,konthasinghe12,kiraz04}, 
and for other quantum technologies~\cite{benjamin09}. 

Thus, under appropriate conditions, 
the emission properties of a driven QD can 
bear close resemblance to the more idealised case of a driven atom in free space. 
QDs are, nevertheless, unavoidably coupled to their 
surrounding solid-state environments.
For coherently-driven (ground state) excitonic transitions in typical arsenide QDs, coupling to acoustic phonons has been demonstrated to dominate the QD-environment interaction~\cite{ramsay10,ramsay10_2}, leading to the appearance of an excitation-induced 
dephasing contribution with a rate that varies with the square of the Rabi frequency (dot-laser coupling strength)~\cite{nazir08,ramsay10,ramsay10_2,ulrich11_short}. 
This driving dependence is theoretically understood as resulting from 
phonons that induce transitions between the dressed states of the QD at the Rabi energy~\cite{machnikowski04,vagov2007_short,mccutcheon10_2,nazir08}, making it the relevant energy scale 
in the three-dimensional phonon environment. 

We shall show here that such transitions can lead to QD emission characteristics that deviate fundamentally 
from the well-established quantum optical behaviour outlined above. 
Specifically, we investigate the competition between photon emission and phonon effects in both the coherent and incoherent scattering properties of a driven QD~\cite{roy11,roy12,ahn05,moelbjerg12,delvalle10}. As our main result, we show that in the presence of phonon coupling the coherent contribution to the QD resonance fluorescence can actually {\it increase} 
with driving strength, in a striking departure from the conventional behaviour in the atomic case. This stems 
from phonon transitions driving thermalisation among the dot dressed states in the system steady-state, 
an effect that arises naturally in our microscopic model of the phonon bath, 
but cannot be captured by a simplified treatment in terms of a  
phenomenological pure dephasing process. As the total scattered light is limited by the photon emission rate, a corresponding {\it decrease} of incoherent emission occurs in the same regime; a trend which a standard quantum optics treatment is again unable to reproduce. 
We also find that, in an appropriate parameter regime, our model 
predicts a narrowing of the Mollow sidebands as the QD-laser detuning is increased, consistent with recent 
experimental observations~\cite{ulrich11_short}. 

We model the QD as a TLS 
with ground state $\ket{0}$ and excited (single exciton) state 
$\ket{X}$, split by an energy $\hbar\w_0$. The dot is 
driven by a laser of frequency $\w_l$, with Rabi frequency $\Omega$, and coupled to two separate harmonic oscillator baths to account for both phonon interactions and spontaneous emission into the radiation field. 
In a frame rotating at frequency $\w_l$, and after a rotating wave approximation on the driving term, 
our Hamiltonian takes the form ($\hbar=1$)  
\begin{equation}\begin{split}
H&=\nu\ketbra{X}{X}+\frac{\Omega}{2}\sigma_x+\sum_\kv \w_\kv b_\kv^{\dagger}b_\kv+\sum_{\bf q}\eta_{\bf q} a_{\bf q}^{\dagger}a_{\bf q}\nonumber\\
&\;+\ketbra{X}{X}\sum_\kv g_\kv(b_\kv^{\dagger}+b_\kv)+\sum_{\bf q} (h_{\bf q} a_{\bf q}\e^{i \w_l t}\sigma_++\mathrm{H.c.}),
\nonumber
\end{split}
\end{equation}
where $\nu=\w_0-\w_l$ is the QD-laser detuning, $\sigma_+=\ketbra{X}{0}$ ($\sigma_-=\sigma_+^{\dagger}$), $\sigma_x=\sigma_+ +\sigma_-$, and ${\mathrm{H.c.}}$ 
denotes the Hermitian conjugate. The 
phonon bath is represented by creation (annihilation) operators $b_\kv^{\dagger}$ ($b_\kv$) for modes with frequency $\w_\kv$, which couple to the QD with strength $g_\kv$. 
The photon bath is similarly defined, with operators $a_{\bf q}^{\dagger}$ ($a_{\bf q}$), frequencies $\eta_{\bf q}$, and couplings $h_{\bf q}$. 

Obtaining an equation of motion for the QD dynamics can be achieved in various ways, 
such as through master equations of weak-coupling~\cite{nazir08,machnikowski04}, polaron~\cite{wilsonrae02,mccutcheon10_2,roy11,roy12}, 
and variational type~\cite{mccutcheon11_2}, as well as by several numerical 
methods~\cite{forstner03,krugel05,vagov2007_short}. For our purposes, master equations are particularly attractive since, with use of the quantum 
regression theorem~\cite{carmichael}, they can readily be applied to investigate emitted field correlation properties~\cite{roy11,roy12}. 
Thus, we opt here to extend 
the variational approach of Ref.~\cite{mccutcheon11_2} 
to include the photon bath, in order 
to calculate field correlations, 
as it is limited neither to weak phonon coupling, nor to 
the small driving limit of polaron theory. 

To the full Hamiltonian we apply a QD-state-dependent phonon displacement 
transformation $H_V=\e^V H \e^{-V}$, with $V=\ketbra{X}{X}\sum_\kv (F(\w_\kv)/\w_\kv)(g_\kv b_\kv^{\dagger}-g_\kv^* b_\kv)$. 
The magnitudes of the displacements are chosen to minimise a free energy bound on the resulting interaction 
terms in $H_V$~\cite{silbey84}. Applying the time-convolutionless projection operator technique to second order in 
the transformed frame, 
we find a master equation of the form~\cite{supplement} 
\beq
\dot{\rho}_{V}=-{\frac{i}{2}}[\epsilon\sigma_z+\Omega_r\sigma_x,\rho_{V}]+\mathcal{K}_{\mathrm{ph}}(\rho_{\rm V})+\mathcal{K}_{\rm{sp}}(\rho_V).
\label{MasterEquation}
\eeq
Here, $\rho_{\rm V}=\mathrm{Tr}_B (\e^V \chi \e^{-V})$, with $\chi$ the complete density operator, is the reduced state 
of the QD TLS in the variational frame, $\epsilon=\nu+\int_0^{\infty}J_{\rm ph}(\omega)\omega^{-1}F(\omega)(F(\omega)-2)\mathrm{d}\w$ and
$\Omega_r=\Omega\exp[-\frac{1}{2}\int_0^{\infty}J_{\rm{ph}}(\w)\w^{-2}F(\w)^2\coth(\beta\w/2)\mathrm{d}\w]$, with temperature $T=1/(k_B\beta)$, are the 
phonon renormalised detuning and Rabi frequency, respectively, 
while $\mathcal{K}_{\rm{sp}}(\rho_V)=\Gamma_1(\sigma_-\rho_{V} \sigma_+-(1/2)\{\sigma_+\sigma_-,\rho_{V}\})$ 
accounts for spontaneous emission. The variational factor $F(\w)=[1-(\epsilon/\xi)\tanh(\beta\xi/2)]
[1-(\epsilon/\xi)\tanh(\beta\xi/2)(1-(\Omega_r^2/2\epsilon\w)\coth(\beta \w/2))]^{-1}$, with $\xi=\sqrt{\epsilon^2+\Omega^2}$, 
is bounded between zero (for no transformation)
and unity (for the polaron transformation), 
while the QD-phonon spectral density 
is usually parameterised by $J_{\rm{ph}}(\w)=\alpha\,\w^3\exp[-(\w/\w_c)^2]$ for coupling to acoustic phonons~\cite{ramsay10,ramsay10_2}. The term $\mathcal{K}_{\mathrm{ph}}(\rho_{\rm V})$, defined in full in the supplementary information, contains all phonon effects other than those included in $\epsilon$ and $\Omega_r$,
representing the various processes induced by phonon interactions, such as pure dephasing, phonon emission, and absorption. 

We characterise the QD photon emission through 
the steady-state first order field correlation 
$g^{(1)}(\tau)=\lim_{t\to\infty}\langle \sigma_+(t)\sigma_-(t+\tau)\rangle$.
The coherent contribution, 
defined as $\smash{g^{(1)}_{\mathrm{coh}}=\lim_{\tau\to\infty}g^{(1)}(\tau)}$, 
is related to the off-diagonal elements of the QD density operator in the steady-state, 
$\smash{g^{(1)}_{\mathrm{coh}}=|\rho_{0X}|^2}$, and is thus a direct consequence of non-vanishing 
QD coherence. The incoherent contribution is then given by $\smash{g^{(1)}_{\mathrm{inc}}(\tau)=g^{(1)}(\tau)-g^{(1)}_{\mathrm{coh}}}$, which determines the incoherent QD emission spectrum via $S_{\mathrm{inc}}(\w)\propto(1/\pi)\mathrm{Re}[\int_0^{\infty}\e^{i (\w-\w_l) \tau}g_{\mathrm{inc}}^{(1)}(\tau)\mathrm{d}\tau]$.

{\it Enhanced coherent scattering.}--We begin our analysis by investigating the emission properties of the QD when driven on resonance with the polaron shifted 
transition frequency ($\epsilon_{F(\omega)\rightarrow1}=0$). We are interested in examining the detailed effects induced by the coupling to phonons as the driving strength is varied. 
In particular, we would like to explore deviations from the 
phenomenological - though often employed and standard in quantum optics~\cite{carmichael} - treatment of 
environmental interactions (beyond radiative decay) 
as giving rise simply to sources of pure dephasing. In fact, we find that the full phonon influence can {\emph{only}} be represented by a pure dephasing form~\cite{supplement}, $\mathcal{K}_{\mathrm{ph}}(\rho_{\rm V})\approx(1/2)\gamma_{\rm PD}(\sigma_z \rho_{\rm V} \sigma_z-\rho_{\rm V})$, for weak resonant driving strengths satisfying $\Omega<k_BT<\omega_c$, consistent with experimental results in this regime~\cite{ramsay10,ulrich11_short,ramsay10_2,muller07,flagg09_short}.
Here, the rate reduces to that given by polaron theory~\cite{roy11,mccutcheon10_2}, 
$\gamma_{\rm PD}=(\Omega_r/2)^2\int_{-\infty}^{\infty}\cos(\Omega_r s)(\e^{\phi(s)}-\e^{-\phi(s)})\mathrm{d}s$, 
where $\phi(s)=\int_0^{\infty}J(\w)\w^{-2}(\cos(\w s)\coth(\beta\w/2)-i\sin(\w s))\mathrm{d}\w$, while $F(\w)\to1$ in Eq.~(\ref{MasterEquation}). Within this limit we can  
derive an analytic expression for $g^{(1)}(\tau)$, giving 
\begin{align}
&g^{(1)}_{\mathrm{inc}}(\tau)=\frac{\Omega_r^2}{2\Omega_r^2+2\Gamma_1\Gamma_{2}}\nonumber\\
\times&\Big[\textstyle{\frac{1}{2}}\e^{-\Gamma_2\tau}+\e^{-\frac{1}{2}(\Gamma_1+\Gamma_2)\tau}(N\cos(\zeta\tau)-M\sin(\zeta\tau))\Big],
\label{g1PureDephasing}
\end{align}
where $\Gamma_2=\frac{1}{2}\Gamma_1+\gamma_{\rm PD}$, $\zeta=\sqrt{\Omega_r^2-(1/4)(\Gamma_1-\Gamma_2)^2}$, 
$N=(\Omega_r^2-\Gamma_1(\Gamma_1-\Gamma_2))/(2\Omega_r^2+2\Gamma_1\Gamma_2)$, and 
$M=(\Omega_r^2(\Gamma_2-3\Gamma_1)+\Gamma_1^3\Gamma_2^2(\Gamma_1^{-1}-\Gamma_2^{-1})^2)/(4\zeta(\Omega_r^2+\Gamma_1\Gamma_2))$, 
and 
\beq
g^{(1)}_{\mathrm{coh}}=\left(\frac{\Gamma_1\Omega_r}{2\Gamma_1\Gamma_2+2\Omega_r^2}\right)^2.
\label{g1PureDephasingCOH}
\eeq
Note that in the pure dephasing model 
$g^{(1)}_{\mathrm{coh}}\rightarrow0$ 
if $\Omega_r$ is allowed to become large, precisely as in the atomic case.

In fact, Eqs.~({\ref{g1PureDephasing}}) and ({\ref{g1PureDephasingCOH}}) are 
essentially the standard atomic 
$g^{(1)}$ expressions when extended to include 
pure dephasing~\cite{flagg09_short,muller07}. The only difference here is that we explicitly include a \emph{driving dependent} pure-dephasing rate, $\gamma_{\rm PD}\sim\Omega_r^2$ (for $\beta\Omega_r,\Omega_r/\w_c\ll 1$), and that the driving is itself renormalised by phonons through $\Omega_r$. 
While both of these features are important 
to approximate the full dynamics, neither will give rise to the kind of pronounced, phonon-induced deviations from standard atomic behaviour in which we are interested. 

\begin{figure}
\begin{center}
\includegraphics[width=0.45\textwidth]{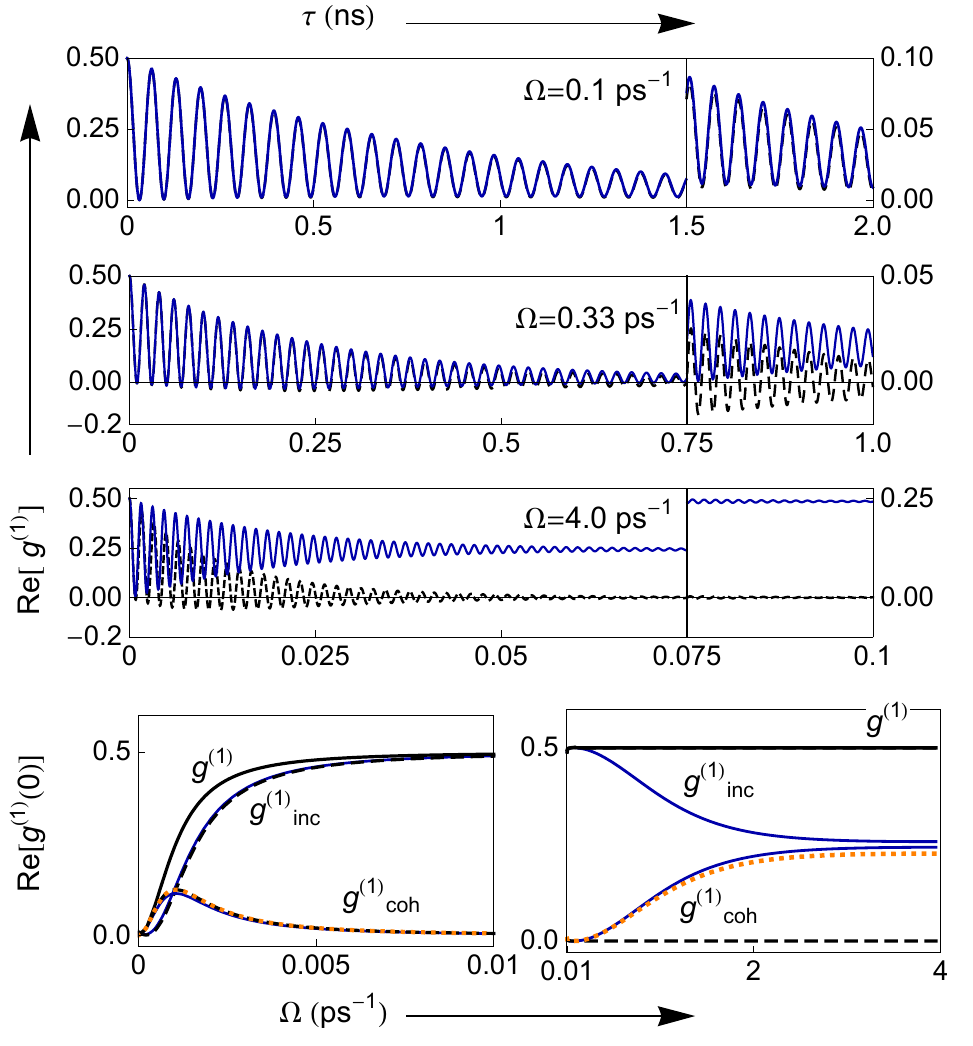}
\caption{Upper three plots: First order field correlation function for various driving strengths, 
as indicated, calculated from the full variational theory (blue solid curves), and the pure dephasing approximation of Eqs.~({\ref{g1PureDephasing}}) and ({\ref{g1PureDephasingCOH}}) (black dashed curves). 
The right-most parts show enlargements of the long-time behaviour. 
Lower plots: Coherent ($g^{(1)}_{\rm coh}$), incoherent ($g^{(1)}_{\rm inc}$), and total ($g^{(1)}$) scattering as a function of driving strength, calculated using the 
full (blue, solid) and pure dephasing (black, dashed) theories.
The total scattering is indistinguishable on this scale between the two models. However, the left plot shows the only region where the pure 
dephasing model gives a non-negligible coherent contribution, 
close to the origin; i.e.,~in the pure dephasing case, all light is incoherently scattered in the right plot. Shown also is $g^{(1)}_{\rm coh}$ calculated from Eq.~(\ref{G1}) (orange dotted curve). Parameters: 
$T_1=700~\mathrm{ps}$, $\alpha=0.027~\mathrm{ps}^2$, $\w_c=2.2~\mathrm{ps}^{-1}$, and $T=4$~K.}
\label{resg1s}
\vspace{-0.5cm}
\end{center}
\end{figure}

To exemplify the breakdown of the pure-dephasing model, in Fig.~{\ref{resg1s}} we plot $g^{(1)}(\tau)$ calculated using the full variational theory (solid blue curves) and 
calculated using Eqs.~({\ref{g1PureDephasing}}) and ({\ref{g1PureDephasingCOH}}) 
(black dashed curves). As expected, for weaker driving, $\Omega<0.1~\mathrm{ps}^{-1}$, the pure dephasing model 
gives a good approximation to the full theory. Nevertheless, as the driving strength is increased, significant discrepancies soon become apparent. In particular, from the 
different long-time values approached when $\Omega\geq0.33~\mathrm{ps}^{-1}$, we 
conclude that the coherent contribution surprisingly 
becomes important in this regime, 
and that this feature is not captured by the pure dephasing approximation. 
Indeed, when $\Omega=4~\mathrm{ps}^{-1}$, 
the full phonon theory gives $\smash{g^{(1)}_{\rm coh}}\sim 0.25$, 
in clear distinction to the pure dephasing case.

That Eqs.~({\ref{g1PureDephasing}}) and ({\ref{g1PureDephasingCOH}}) cannot capture these effects signifies that above a driving strength of
$\Omega\sim0.1~\mathrm{ps}^{-1}$ (for these realistic parameters), the field correlation properties of the QD emission 
fundamentally depart from the atomic case. At driving above saturation, photons mediate transitions between manifolds of the dot-laser dressed states, while phonons mediate transitions between dressed states in a single manifold. Hence, photon emission acts in this regime to completely suppress QD coherences in the steady-state, while phonons drive thermalisation among the dressed states, thus leading to QD steady-states with non-negligible coherence. 
When phonon processes dominate over photon emission, as in the strong-driving regime, we then find that the level of coherent emission correspondingly grows. 
Though the pure dephasing model correctly captures the fact that phonon-induced damping remains driving-dependent across the full parameter range, it fails here because it does not lead to the correct equilibration of the QD with 
the phonon bath. In this regard, it assumes a high temperature limit with respect to the driving strength, 
and thus the quantum nature of the environment is lost. 

For resonant driving, we can (approximately) rectify this by the modification $\mathcal{K}_{\mathrm{ph}}(\rho_{\rm V})\approx(1/2)\gamma_{\rm PD}(\sigma_z \rho_{\rm V} \sigma_z-\rho_{\rm V})+(i/4)\kappa[\sigma_y,\{\sigma_z,\rho_{\rm V}\}]$, 
where $\kappa=(\Omega_r/2)^2\int_{-\infty}^{\infty}\sin(\Omega_r s)(\e^{\phi(s)}-\e^{-\phi(s)})\mathrm{d}s$, such that $\kappa/\gamma_{\rm PD}=\tanh(\beta\Omega_r/2)$. 
We now find
\beq
g^{(1)}_{\mathrm{coh}}\rightarrow G^{(1)}_{\mathrm{coh}}=\left(\frac{\Gamma_1\Omega_r}{2\Gamma_1\Gamma_2+2\Omega_r^2}\right)^2
+\left(\frac{\Omega_r\kappa/\Omega}{\Gamma_1+2\gamma_{\rm PD}}\right)^2,
\label{G1}
\eeq
where the first term is precisely the contribution in the strict pure-dephasing case [see Eq.~({\ref{g1PureDephasingCOH}})], 
which quickly becomes negligible for large 
$\Omega_r$. Conversely, the second term, 
now arising due to equilibration with the 
quantum mechanical phonon bath, becomes important as $\Omega_r$ increases. 
To see this, we note that once $\Omega_r$ is large enough such that 
$\Gamma_1\ll\gamma_{\rm PD}$, 
we can approximate $G^{(1)}_{\mathrm{coh}}\approx(\Omega_r\tanh(\beta\Omega_r/2)/2\Omega)^2$. 
In the upper three plots of Fig.~{\ref{resg1s}}, increasing the driving moves the QD from an effective 
high temperature regime, where $\beta\Omega_r\ll1$ and $G^{(1)}_{\mathrm{coh}}\approx 0$, to an effective low 
temperature regime, where $\beta\Omega_r\gg1$ and $G^{(1)}_{\mathrm{coh}}\approx (\Omega_r/2\Omega)^2$. 
These observations are borne out in 
the lower part of Fig.~{\ref{resg1s}}, where we plot the coherent, incoherent, and total scattering as a function of $\Omega$. 
The lower left plot shows the region close to the origin, the {\emph{only regime}} in which the pure dephasing model predicts a non-negligible level of coherent emission.
From the lower right plot, we see also that as the total scattering is fixed at strong driving, the incoherent contribution decreases in our full phonon model as the coherent contribution increases. Again, this is not captured by the pure dephasing treatment. In fact, this represents a hitherto unexplored regime of resonance fluorescence at strong driving, in which both significant coherent scattering and a well-defined Mollow triplet can coexist.

QD resonance fluorescence experiments are usually performed at Rabi frequencies up to around $25$ GHz
($\Omega=0.16$~ps$^{-1}$), at which point the coherent fraction is of order 
$1$ \% from our full phonon model, compared to $0.01$ \% in the pure dephasing model (for the parameters of Fig.~\ref{resg1s}). Increasing $\Omega$ fourfold, around 15 \% of the light is then coherently scattered in the full model, compared to less than 0.0003 \% in the pure dephasing case. Reducing the temperature to $2$~K, the coherent fraction could be increased to about 35 \% at this driving strength. 

{\it Spectrum.}-- 
We now turn our attention to the 
QD emission spectrum, concentrating on cases where the incoherent contribution dominates (i.e.~relatively weak driving), 
and Eq.~(\ref{g1PureDephasing}) is thus approximately valid on resonance. 
From a Fourier transform of Eq.~(\ref{g1PureDephasing}), we find that the resonant Mollow sideband 
widths are determined by $\Gamma_1+\Gamma_2=(3/2)\Gamma_1+\gamma_{\rm PD}$, 
with approximate positions $\pm\Omega_r$. We therefore expect a systematic broadening and splitting with increasing driving strength~\cite{ulrich11_short, roy11}. 
Off resonance, we might then also expect sideband broadening and splitting with increasing {\emph{detuning}} $\epsilon$ (for fixed $\Omega$) if we were to replace $\Omega_r$ with the generalised Rabi frequency, $\Omega_r'=\sqrt{\Omega_r^2+\epsilon^2}$~\cite{ulrich11_short}, leading to similar trends for increasing $\epsilon$ as for $\Omega$. 
However, the experiments of Ref.~\cite{ulrich11_short} showed a systematic {\emph{narrowing}} of the Mollow sidebands with increasing detuning, 
leaving open the question as to why this might be the case.  

\begin{figure}
\begin{center}
\includegraphics[width=0.43\textwidth]{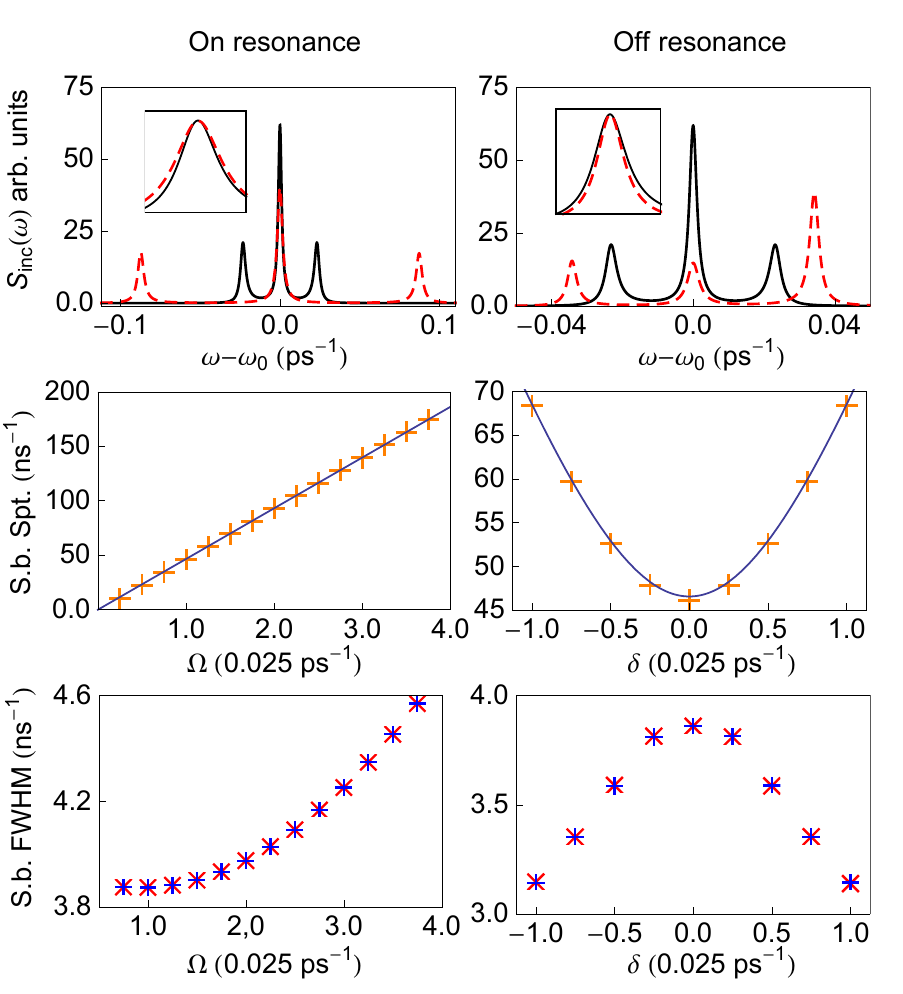}
\caption{From top to bottom, incoherent emission spectrum, extracted sideband splitting, 
and extracted sideband width for varying driving strength on resonance (left), and 
varying detuning (right). The solid black curves in the emission spectra are for $\epsilon=0$, 
and a driving strength of $\Omega=0.025~\mathrm{ps}^{-1}$ (which sets our x-axis units in the rest of the plots). The dashed red curves are for $\Omega=0.094~\mathrm{ps}^{-1}$ on resonance, 
and $\epsilon=\Omega=0.025~\mathrm{ps}^{-1}$ off resonance (which has been enhanced by a factor of $5$). The insets show the red sidebands shifted and rescaled to lie on top of each other. The solid blue curves in the middle row show the functions $2\Omega_r$ (left) and $2\sqrt{\Omega_r^2+\epsilon^2}$ (right). The symbols in the bottom row correspond to the red ($\times$) and blue ($+$) sidebands. Parameters: $T_1=400~\mathrm{ps}$, $\alpha=0.027~\mathrm{ps}^2$, $\w_c=2.2~\mathrm{ps}^{-1}$, and $T=10$~K.}
\label{ORSpec}
\vspace{-0.5cm}
\end{center}
\end{figure}

In fact, off-resonance the expressions for the spectrum become significantly more complicated than 
in the resonant case, and the above simple reasoning does not hold. 
To illustrate this, in Fig.~{\ref{ORSpec}}, from top to bottom, we plot the incoherent emission spectrum, extracted sideband splitting, and extracted 
full-width-half-maximum ($=\Gamma$) of the Mollow sidebands, calculated from the full phonon theory. 
In the latter two cases, the spectrum is fitted 
by a sum of three 
Lorentzian functions of the form $L(\w)= 0.5\Gamma/[(\w-\w_p)^2+(0.5\Gamma)^2]$. The left column corresponds to varying the 
driving frequency on resonance, while the right column corresponds to varying the detuning with a fixed driving strength. 

As can be seen by the sideband splittings in the middle left plot, increasing the 
driving strength on resonance does, as expected, cause the sidebands to move apart linearly with $\Omega$. 
Also, we see in the middle right plot that moving off-resonance 
appears to alter the sideband splitting in exact accordance with the simple procedure of replacing 
$\Omega_r\to\sqrt{\Omega_r^2+\epsilon^2}$. The extracted sideband widths in the lower plots, however, reveal something quite different. On resonance, in accordance with $\gamma_{\rm PD}\sim\Omega_r^2$, we see a systematic broadening of the sidebands with increasing driving strength. 
In contrast, as we move off-resonance, we now see a systematic 
\emph{narrowing} of the sidebands, consistent with recent experimental results~\cite{ulrich11_short}. To further confirm this point, the insets of the 
plots in the top row show the red sidebands in each case plotted on top of each other. 
We can gain some approximate analytical insight into this behaviour for small detuning by again considering the pure dephasing limit. 
Allowing for off-resonant driving, we expand the sideband widths to second order in the detuning, from which we find that they are determined by $(3/2)\Gamma_1+\gamma_{\rm PD}-(\epsilon/\sqrt{2}\Omega_r)^2(\Gamma_1-2\gamma_{\rm PD})$. Hence, for $\Gamma_1>2\gamma_{\rm PD}$, as in Fig.~\ref{ORSpec}, we expect narrowing as we detune 
from resonance, while broadening occurs for $\Gamma_1<2\gamma_{\rm PD}$. Note 
that while we do not include detailed cavity effects here, 
which give rise to qualitatively different behaviour in Refs.~\cite{roy11,roy12}, 
our results demonstrate that for a QD TLS at least, an increase in sideband splitting off-resonance 
does not necessarily imply an associated phonon-induced increase in sideband width.

{\it Summary.}--
We have shown that the balance of coherent to incoherent emission from a driven TLS can be fundamentally altered by environmental interactions, 
leading to a nonstandard regime of resonance fluorescence attainable in solid-state emitters. In the context of driven QDs, enhanced coherent scattering can occur with increasing driving strength, due to thermalisation in the QD steady-state with respect to the phonon bath. This mechanism is in fact 
rather general, 
and could occur for any emitter 
in which the steady-state becomes dominated by dressed state thermalisation. For off-resonant driving, we have shown that QD-phonon interactions do not necessarily lead to broadening in the spectral sideband widths with increasing detuning. 
In fact, narrowing can occur in certain regimes, consistent with an observed experimental trend~\cite{ulrich11_short}. 
Again, this behaviour is not QD-specific, and so we expect the emission features outlined above to be of importance in a wide variety of experimental settings.

{\it Acknowledgments} - 
During the completion of this work 
we became aware of similar results for the spectral narrowing obtained independently~\cite{ulhaq13}. We thank Stephen Hughes and co-workers for bringing these to our attention. We also thank Clemens Matthiesen, Brendon Lovett, Erik Gauger, and Sean Barrett for fruitful discussions. D.P.S.M. acknowledges support from the EPSRC, CHIST-ERA project SSQN, and CONICET. A.N. is supported by Imperial College.

\clearpage

\widetext
\section{Supplemental Material}

In this supplement we outline the derivation of the master equation used in the main text, Eq.~(1). We first show how the quantum dot-phonon and quantum dot-photon coupling effects can be treated independently within our formalism. We then give expressions from the variational method used to treat the quantum dot-phonon coupling, 
and show how they can be approximated by a pure dephasing form in the appropriate (weak-driving) limit.

\subsection{Separation of phonon and photon terms}

Our starting point is the quantum dot (QD) Hamiltonian as given in the main text:
\begin{eqnarray}\label{HRWA}
H=\nu|X\rangle\langle X|+\frac{\Omega}{2}\sigma_x+\sum_{\bf k}\omega_{\bf k}b_{\bf k}^{\dagger}b_{\bf k}+
|X\rangle\langle X|\sum_{\bf k}g_{\bf k}(b_{\bf k}^{\dagger}+b_{\bf k})+\sum_{\bf q}\eta_{\bf q}a_{\bf q}^{\dagger}a_{\bf q}
+\sum_{\bf q}(h_{\bf q}a_{\bf q}\sigma_+e^{i\omega_lt}+{h^*_{\bf q}}a_{\bf q}^{\dagger}\sigma_-e^{-i\omega_lt}).
\end{eqnarray}
Following Ref.~\cite{mccutcheon11_2}, we first apply a unitary variational transformation in order to 
treat the QD-phonon interaction beyond the weak coupling approximation. 
The transformed Hamiltonian is defined by $H_{V}=e^{V}He^{-V}$, where 
\begin{equation}
\label{poltrans}
\exp[\pm V]=\exp\bigg[\pm |X\rangle\langle X|\sum_{\bf k}(\alpha_{\bf k}b_{\bf k}^{\dagger}-\alpha_{\bf k}^*b_{\bf k})\bigg]=|0\rangle\langle0|+|X\rangle\langle X|\prod_{\bf k}D(\pm\alpha_{\bf k}),
\end{equation}
with $D(\pm\alpha_{\bf k})=\exp[\pm(\alpha_{\bf k}b_{\bf k}^{\dagger}-\alpha_{\bf k}^*b_{\bf k})]$ and $\alpha_{\bf k}=f_{\bf k}/\omega_{\bf k}$. 
Here, $f_{\bf k}$ are variational parameters to be determined later. 
After the transformation, we write $H_V=H_S+H_{1a}+H_{1b}+H_{12}+H_{B_1}+H_{B_2}$, where
\beq
H_S=\frac{R}{2}\openone+\frac{\epsilon}{2}\sigma_z+\frac{\Omega_r}{2}\sigma_x,
\eeq
and the interaction terms $H_{1a}=|X\rangle\langle X|\sum_{\bf k}(g_{\bf k}-f_{\bf k})(b_{\bf k}^{\dagger}+b_{\bf k})$ 
and $H_{1b}=\frac{\Omega}{2}\left(\sigma_xB_x+\sigma_yB_y\right)$, with 
$B_x=\frac{1}{2}(B_++B_--2B)$ and $B_y=\frac{1}{2i}(B_--B_+)$ for $B_{\pm}=\prod_{\bf k}D(\pm\alpha_{\bf k})$, contain only QD and phonon operators. The interaction term
\beq
H_{12}=\sum_{\bf q}(h_{\bf q}a_{\bf q}B_+\sigma_+e^{i\omega_lt}+{h^*_{\bf q}}a_{\bf q}^{\dagger}B_-\sigma_-e^{-i\omega_lt}),
\label{H2}
\eeq
contains QD, phonon, and photon operators, 
and the bath Hamiltonians are $H_{B_1}=\sum_{\bf k}\omega_{\bf k}b_{\bf k}^{\dagger}b_{\bf k}$ and $H_{B_2}=\sum_{\bf q}\eta_{\bf q}a_{\bf q}^{\dagger}a_{\bf q}$. 
The detuning now becomes $\epsilon=\omega_0'-\omega_l$, defined in terms of the {\it bath-shifted} QD transition energy $\omega_0'=\omega_0+R$, with $R=\sum_{\bf k}\omega_{\bf k}^{-1}f_{\bf k}(f_{\bf k}-2g_{\bf k})$. 
We assume a thermal equilibrium state for the phonon and photon baths, 
$\rho_{B}=e^{-\beta H_{B}}/{\rm tr}_B(e^{-\beta H_{B}})=[e^{-\beta H_{B_1}}/{\rm tr}_{B_1}(e^{-\beta H_{B_1}})][e^{-\beta H_{B_2}}/{\rm tr}_{B_2}(e^{-\beta H_{B_2}})]=\rho_{B_1}\rho_{B_2}$, 
and in doing so find that the operators $B_{\pm}$ have the same average with respect to this state: $B=\rm{tr}( B_{\pm}\rho_B)=\exp[-(1/2)\sum_{\bf k}|\alpha_{\bf k}|^2\coth{(\beta\omega_{\bf k}/2)}]$, 
with inverse temperature $\beta=1/(k_BT)$. The {\it bath-renormalised} Rabi frequency is defined as $\Omega_r=B\Omega$.

We now separate the variationally-transformed Hamiltonian into 
$H_V=H_0+H_I$, 
with $H_0=H_S+H_{B_1}+H_{B_2}$ and $H_I=H_{1a}+H_{1b}+H_{12}$, and treat $H_I$ as a perturbation. 
We move into the interaction picture with respect to $H_0$, yielding an interaction Hamiltonian in the (variationally-transformed) interaction picture 
of the form $\tilde{H}_I(t)=\tilde{H}_{1a}(t)+\tilde{H}_{1b}(t)+\tilde{H}_{12}(t)$, where $\tilde{H}_{1a}(t)=e^{iH_0t}H_{1a}e^{-iH_0t}$, $\tilde{H}_{1b}(t)=e^{iH_0t}H_{1b}e^{-iH_0t}$, 
and 
\begin{equation}
\label{Hint2}
\tilde{H}_{12}(t)=e^{iH_0t}H_{12}e^{-iH_0t}=\sum_{\bf q}(h_{\bf q}a_{\bf q}e^{-i\eta_{\bf q}t}B_+(t)\sigma_+(t)e^{i\omega_lt}+{h^*_{\bf q}}a_{\bf q}^{\dagger}e^{i\eta_{\bf q}t}B_-(t)\sigma_-(t)e^{-i\omega_lt}).
\end{equation}
Here, $B_{\pm}(t)=e^{iH_{B_1}t}B_{\pm}e^{-iH_{B_1}t}=\prod_{\bf k}D(\pm\alpha_{\bf k}e^{i\omega_{\bf k}t})$, and 
\beq
\sigma_{\pm}(t)e^{\pm i\omega_lt}=\exp\Big[i\Big(\frac{\epsilon}{2}\sigma_z+\frac{\Omega_r}{2}\sigma_x\Big)t\Big]\sigma_{\pm}\exp\Big[-i\Big(\frac{\epsilon}{2}\sigma_z+\frac{\Omega_r}{2}\sigma_x\Big)t\Big]e^{\pm i(\omega_0-\nu)t}.
\eeq
Provided $\omega_0\gg\nu,\epsilon,\Omega_r$, which is generally the case for driven QDs since $\omega_0\sim1$~eV compared to meV or smaller energy scales for the other quantities, 
we can then approximate $\sigma_{\pm}(t)e^{\pm i\omega_lt}\approx\sigma_{\pm}e^{\pm i\omega_0t}$. 

Following the standard projection-operator procedure we derive a time-local master equation 
for the reduced QD exciton density operator, $\tilde{\rho}_V$, in the variational frame interaction picture. 
Choosing the bath reference state to be $\rho_B$ used above, we find
\begin{equation}
\label{tcl2}
\frac{\rm{d}\tilde{\rho}_{V}(t)}{\rm{d}t}=-\int_0^t\mathrm{d}s{\rm tr}_B[\tilde{H}_{I}(t),[\tilde{H}_{I}(s),\tilde{\rho}_{V}(t)\rho_{B}]].
\end{equation}
Since ${\rm{tr}}_{B_1}(\tilde{H}_{1a}(t)\rho_{B_1})={\rm{tr}}_{B_1}(\tilde{H}_{1b}(t)\rho_{B_1})={\rm{tr}}_{B_2}(\tilde{H}_{12}(t)\rho_{B_2})=0$,
we find that Eq.~({\ref{tcl2}}) can be written 
$\frac{d\tilde{\rho}_{V}(t)}{dt}=\tilde{\mathcal{K}}_{\mathrm{ph}}(\tilde{\rho}_V(t))+\tilde{\mathcal{K}}_{\mathrm{sp}}(\tilde{\rho}_V(t))$, where
\begin{equation}
\label{phononterm}
\tilde{\mathcal{K}}_{\mathrm{ph}}(\tilde{\rho}_V(t))=-\int_0^t\mathrm{d}s{\rm tr}_{B_1}[\tilde{H}_{1a}(t)+\tilde{H}_{1b}(t),[\tilde{H}_{1a}(s)+\tilde{H}_{1b}(s),\tilde{\rho}_{V}(t)\rho_{B_1}]]
\end{equation}
and is precisely the form expected from the variational treatment of QD exciton-phonon interactions in the absence of the radiation field, whereas 
\begin{equation}
\label{radiationterm}
\tilde{\mathcal{K}}_{\mathrm{sp}}(\tilde{\rho}_V(t))=-\int_0^t\mathrm{d}s{\rm tr}_{B_1+B_2}[\tilde{H}_{12}(t),[\tilde{H}_{12}(s),\tilde{\rho}_{SP}(t)\rho_{B_1}\rho_{B_2}]],
\end{equation}
is responsible for photon emission and absorption processes. Though the latter term appears at this stage to be modified by the phonon environment 
due to our use of the variational transformation, we shall now show that the modification is negligible for the situation considered in this work. 

\subsection{Spontaneous emission terms}

To proceed, we write $\tilde{H}_{12}(t)=A(t)Q(t)B_+(t)+(A(t)Q(t)B_+(t))^{\dagger}$, 
with $A(t)=\sigma_+e^{i\omega_0t}$ and $Q(t)=\sum_{\bf q}h_{\bf q}a_{\bf q}e^{-i\eta_{\bf q}t}$, where $A(t)$, $Q(t)$, and $B_+(t)$ all 
commute. Inserting this into Eq.~(\ref{radiationterm}), 
we find that the radiation field term can be written in the simple and familiar form 
\begin{equation}
\label{radiationmasterterms}
\tilde{\mathcal{K}}_{\mathrm{sp}}(\tilde{\rho}_V(t))=\Gamma_1(t)\left(\sigma_-\tilde{\rho}_{V}(t)\sigma_+-\frac{1}{2}\{\sigma_+\sigma_-,\tilde{\rho}_{V}(t)\}\right).
\end{equation}
where we ignore absorption and stimulated emission processes under the assumption that no thermal photons exist at the appropriate energy scale 
for temperatures of interest. Additionally, we have ignored the Lamb-shift of the excitonic energy splitting induced by the radiation field. 
The rate of spontaneous emission processes is given by
\begin{equation} 
\Gamma_1(t)=2{\rm Re}\int_0^t\mathrm{d}\tau e^{i\omega_0\tau}C(\tau)X(\tau),
\label{sprate2}
\end{equation}
where in the continuum limit of the phonon bath
\begin{equation}
\label{phononcorr}
C(\tau)={\rm{tr}}_{B_1}(B_{\pm}(\tau)B_{\mp})=\exp{\bigg[-\int_0^{\infty}\mathrm{d}\omega \frac{J_{\mathrm{ph}}(\omega)}{\omega^2}F(\w)^2((1-\cos{\omega\tau})\coth{\beta\omega/2}+i\sin{\omega\tau})\bigg]},
\end{equation}
with $f_{\bf k}/g_{\bf k}=F(\w_{\bf k})$ (which will be justified later), and $J_{\mathrm{ph}}(\omega)=\sum_{\bf k}|g_{\bf k}|^2\delta(\omega-\omega_{\bf k})$ is the phonon spectral density. In the continuum limit of the photon bath  
$X(\tau)=\int_0^{\infty}\mathrm{d}\eta e^{-i\eta\tau}J_{\mathrm{pt}}(\eta)$, 
where $J_{\mathrm{pt}}(\eta)=\sum_{\bf q}|h_{\bf q}|^2\delta(\eta-\eta_{\bf q})$ is the relevant photon spectral density. Thus, the spontaneous emission rate we derive within the variational theory is dependent upon both the phonon and photon bath correlation functions, $C(\tau)$ and $X(\tau)$, respectively, 
and whether this rate varies from that in the absence of the phonon environment depends crucially on their respective timescales.

We know that the typical timescale for the phonon-bath correlation function to reach the 
long-time value of $B^2$ is of the order of a few picoseconds~\cite{ramsay10,ramsay10_2}. For $X(\tau)$, we take the standard (3D) spectral 
density $J_{\rm{pt}}(\eta)=A\,\eta^3e^{-\eta/\eta_c}$, where a high-frequency cut-off $\eta_c$ has been introduced. This gives 
\begin{equation}\label{Xcorr}
X(\tau)=\frac{6A}{(\eta_c^{-1}+i\tau)^4},
\end{equation}
which decays to zero on a timescale of roughly $1/\eta_c$. 
For spontaneous emission not to be suppressed, it must be the case that $\eta_c>\w_0$.
Thus, we can estimate $\eta_c>1.5\times10^3$~ps$^{-1}$, for a typical $|0\rangle$ to $|X\rangle$ energy splitting of $1$~eV, which leads to a radiation field correlation time of the order of 
femtoseconds {\it at most}. On this timescale, 
the phonon correlation function barely changes, and we may replace $C(\tau)$ by $C(0)=1$ in Eq.~(\ref{sprate2}), and we are also now justified in 
taking the upper limit of integration to infinity for timescales of interest. Thus, for a typical QD system as described in the main text, the spontaneous emission process is unaltered by the exciton-phonon coupling, 
and can be described by the standard Lindblad form of Eq.~({\ref{radiationmasterterms}}) with 
$\Gamma_1(t)$ replaced with $\Gamma_1=2\pi J_{\rm{pt}}(\omega_0)$. 

\subsection{Phonon coupling terms}

We now use the methods described in Ref.~\cite{mccutcheon11_2} 
to find the form of the phonon terms. The variational parameters upon which $H_{1a}$, $H_{1b}$ and $H_S$ all depend 
are found by minimising a free energy bound on the interaction terms. 
We find $f_{\bf k}=g_{\bf k}F(\w_{\bf k})$ with
\begin{align}
F(\w_\kv)=\frac{(1-\frac{\epsilon}{\xi}\tanh(\beta\xi/2))}
{1-\frac{\epsilon}{\xi}\tanh(\beta\xi/2)\Big(1-\frac{\Omega_r^2}{2\epsilon\w_\kv}\coth(\beta \w_\kv/2)\Big)},
\label{minimisation_condition}
\end{align}
and $\xi=\sqrt{\Omega^2+\epsilon^2}$. We note that since $\Omega_r=\Omega B$ and $\epsilon=\nu+R$ are functions of $F(\w_\kv)$ their values 
must be solved for self-consistently.

Moving the phonon coupling terms back into the Schr\"{o}dinger picture, $\mathcal{K}_{\rm{ph}}(\rho_V(t))=\e^{-i H_0 t}\tilde{\mathcal{K}}_{\mathrm{ph}}(\tilde{\rho}_V(t))\e^{i H_0 t}$, we find
\begin{align}
\mathcal{K}_{\rm{ph}}(\rho_V(t))=&-{\textstyle{\frac{1}{2}}}\sum_{ij}\sum_{\w}\gamma_{ij}(\w)[A_i,A_j(\w)\rho_V(t)-\rho_V(t)A_j^{\dagger}(\w)]\nonumber\\
&-i\sum_{ij}\sum_{\w}S_{ij}(\w)[A_i,A_j(\w)\rho_V(t)+\rho_V(t)A_j^{\dagger}(\w)],
\label{me_2}
\end{align}
where $\{i, j\}\in\{1,2,3\}$ and $\omega\in\{0,\pm\xi\}$. We define $A_1=\sigma_x$, $A_2=\sigma_y$ and $A_3=(1/2)(I+\sigma_z)$, while 
$A_1(0)=\sin 2\theta(\ketbra{+}{+}-\ketbra{-}{-})$, $A_1(\xi)=\cos 2\theta\ketbra{-}{+}$, 
$A_2(0)=0$, $A_2(\xi)=i \ketbra{-}{+}$, $A_3(0)=\cos^2\theta\ketbra{+}{+}+\sin^2\theta\ketbra{-}{-}$ and 
$A_3(\xi)=-\sin\theta\cos\theta\ketbra{-}{+}$, defined in terms of the eigenstates of $H_S$, satisfying 
$H_S\ket{\pm}=(1/2)(R\pm\xi)\ket{\pm}$. In all cases $A_i(\w)=A_i^{\dagger}(-\w)$, 
and $\theta=(1/2)\arctan(\Omega_r/\epsilon)$. Eq.~({\ref{me_2}}) contains the quantities
$\gamma_{ij}(\w)=2\mathrm{Re}[K_{ij}(\w)]$ and $S_{ij}(\w)=\mathrm{Im}[K_{ij}(\w)]$, defined in terms of the response functions
\beq\label{bathresponse}
K_{ij}(\w)=\int_0^{\infty} \Lambda_{ij}(\tau)\e^{i\w t}\mathrm{d}\tau,
\eeq
which themselves depend on the bath correlation functions $\Lambda_{ij}(\tau)=\mathrm{tr}(\tilde{B}_i(\tau)\tilde{B}_j(0)\rho_B)$. Note that in 
Eq.~({\ref{bathresponse}}) we have extended the upper limit of integration to infinity which, for the parameters considered in the main text, is a good approximation~\cite{mccutcheon10_2}. We label the bath operators 
$B_1=(\Omega/2)B_x$, $B_2=(\Omega/2)B_y$, and $B_3=\sum_\kv(g_\kv-f_\kv)(b_\kv^{\dagger}+b_\kv)$. The bath correlation functions are found to be 
$\Lambda_{11}(\tau)=(\Omega_r^2/8)(\e^{\phi(\tau)}+\e^{-\phi(\tau)}-2)$ and $\Lambda_{22}(\tau)=(\Omega_r^2/8)(\e^{\phi(\tau)}-\e^{-\phi(\tau)})$, 
with phonon propagator 
\begin{equation}
\phi(\tau)=\int_0^{\infty}\mathrm{d}\w\frac{J(\w)}{\w^2}F(\w)^2G_+(\tau),
\eeq
defined in terms of $G_{\pm}(\tau)=(n(\w)+1)e^{-i\w\tau}\pm n(\w)\e^{i\w\tau}$, with $n(\w)=(e^{\beta \w}-1)^{-1}$ the occupation number, while
\beq
\Lambda_{33}(\tau)=\int_0^{\infty}\mathrm{d}\w J(\w)(1-F(\w))^2G_+(\tau),
\qquad
\Lambda_{32}(\tau)=\frac{\Omega_r}{2}\int_0^{\infty}\mathrm{d}\w\frac{J(\w)}{\w}F(\w)(1-F(\w))iG_-(\tau),
\eeq
with $\Lambda_{32}(\tau)=-\Lambda_{23}(\tau)$, and $\Lambda_{12}(\tau)=\Lambda_{21}(\tau)=\Lambda_{13}(\tau)=\Lambda_{31}(\tau)=0$.

Putting everything together, we arrive at the full variational frame Schr\"{o}dinger picture master equation
\beq
\frac{\rm{d}\rho_V(t)}{\rm{d}t}=-\frac{i}{2}[\epsilon\sigma_z+\Omega_r\sigma_x,\rho_V(t)]+\mathcal{K}_{\rm{ph}}(\rho_V(t))
+\Gamma_1\left(\sigma_-\rho_{V}(t)\sigma_+-\frac{1}{2}\{\sigma_+\sigma_-,\rho_{V}(t)\}\right),
\label{FinalME}
\eeq
as used in the main text. We note that in moving the spontaneous emission terms back into the Schr\"{o}dinger picture the QD 
operators $\sigma_{\pm}$ have remained unchanged to be consistent with the approximation that 
$\sigma_{\pm}(t)e^{\pm i\omega_lt}\approx\sigma_{\pm}e^{\pm i\omega_0t}$ used previously.

\subsection{Pure dephasing limit}

Though we use the full form of Eq.~({\ref{me_2}}) for the phonon coupling in numerically calculating the field correlation properties of the QD, the analytical expressions resulting from it are somewhat cumbersome. However, 
in the correct (weak-driving) limit, we find that the phonon coupling terms in Eq.~({\ref{me_2}}) can be well approximated by a simple pure dephasing form. Assuming now that 
we drive the QD on resonance with the polaron shifted transition frequency, $\nu-\sum_\kv \w^{-1}g_\kv^2=0$, and we drive weakly enough such that $\Omega\ll\omega_c$, then the variational transformation reduces approximately to the full polaron form, and we can thus set $F(\w_\kv)\approx1$ in Eq.~({\ref{minimisation_condition}}). As such, we find that only the correlation functions $\Lambda_{11}(\tau)$ and $\Lambda_{22}(\tau)$ survive, and the variational master equation (now ignoring spontaneous emission) reduces to the polaron form given in Ref.~\cite{mccutcheon10_2}. 
This corresponds to Bloch equations of the form $\dot{\vec{\alpha}}=M\cdot\vec{\alpha}+\vec{b}$,
where
\beq
M=\left(
\begin{array}{ccc}
-(\Gamma_z-\Gamma_y) & 0 & 0 \\
0 & -\Gamma_y & -\Omega_r \\
0 & (\Omega_r+\lambda) & -\Gamma_z
\end{array}
\right),
\eeq
and $\vec{b}=(-\kappa_x,0,0)^T$, with Bloch vector $\vec{\alpha}=(\langle\sigma_x \rangle_{t}, \langle\sigma_y \rangle_{t}, \langle\sigma_z \rangle_{t})^T$. 
The rates and energy shifts are given by $\Gamma_y=2\gamma_{11}(0)$, $\Gamma_z=\gamma_{22}(\Omega_r)+\gamma_{22}(-\Omega_r)$, 
$\kappa=\gamma_{22}(\Omega_r)-\gamma_{22}(-\Omega_r)$, and $\lambda=2[S_{22}(\Omega_r)-S_{22}(-\Omega_r)]$.
We are interested in the solutions to these Bloch equations for arbitrary initial conditions as, with the help of the regression theorem, this will determine the phonon contribution to the first-order field correlation function, and hence the QD emission spectrum. For the appropriate phonon spectral density used in the main text, $J_{\rm{ph}}(\w)=\alpha\,\w^3\exp[-(\w/\w_c)^2]$, we find that in the regime that $k_BT<\omega_c$, 
as the driving strength becomes small, then $\gamma_{11}(0)$ becomes negligible in comparison to $\gamma_{22}(\pm\Omega_r)$. 
Hence, $\Gamma_{y}\rightarrow0$, while $\Gamma_z\rightarrow(\gamma_{22}(\Omega_r)+\gamma_{22}(-\Omega_r))$. Additionally, $\lambda\ll\Omega_r$ in this regime, 
and $(\Gamma_z-\Gamma_y)/\zeta\approx\Gamma_z/\sqrt{\Omega_r^2-(1/4)\Gamma_z^2}$ is very small as well. If we additionally impose $\Omega\beta\ll1$, such that $\kappa/\Gamma_z\rightarrow0$, then we may approximate the Bloch equation solutions as 
\begin{eqnarray}
\langle\sigma_x\rangle_t&{}\approx{}&e^{-\Gamma_zt}\langle\sigma_x\rangle_0,\label{polsolspuresx}\\
\langle\sigma_y\rangle_t&{}\approx{}&e^{-\Gamma_zt/2}\left[\langle\sigma_y\rangle_0\cos{(\zeta t)}-\frac{\Omega_r}{\zeta}\langle\sigma_z\rangle_0\sin{(\zeta t)}\right],\label{polsolspuresy}\\
\langle\sigma_z\rangle_t&{}\approx{}&e^{-\Gamma_zt/2}\left[\langle\sigma_z\rangle_0\cos{(\zeta t)}+\frac{\Omega_r}{\zeta}\langle\sigma_y\rangle_0\sin{(\zeta t)}\right],\label{polsolspuresz}
\end{eqnarray}
where $\zeta\rightarrow\sqrt{\Omega_r^2-(1/4)\Gamma_z^2}$. Now, if we identify $\gamma_{PD}=\Gamma_z$, then these are precisely the solutions we 
expect from a simple pure dephasing master equation of the form 
\begin{equation}\label{puredephsingmaster}
\dot{\rho}=-\frac{i\Omega_r}{2}[\sigma_x,\rho]+\frac{\gamma_{PD}}{2}(\sigma_z\rho\sigma_z-\rho),
\end{equation}
in the relevant regime of $\gamma_{PD}/\sqrt{\Omega_r^2-(1/4)\gamma_{PD}^2}$ being small. Furthermore, if we wish to ensure that the system tends to the correct steady state in the long-time limit, we then need to add a term $(i\kappa/4)[\sigma_y,\{\sigma_z,\rho\}]$ to the right-hand-side of Eq.~(\ref{puredephsingmaster}), such that the solution for $\langle\sigma_x\rangle_t$ becomes $\langle\sigma_x\rangle_t=\frac{e^{-\Gamma_zt}}{\Gamma_z}\left[\langle\sigma_x\rangle_0\Gamma_z+\kappa\right]-\frac{\kappa}{\Gamma_z}$. This justifies the forms given in the main text for weak driving, and is further confirmed by 
the agreement we see with the full numerical solution of the variational master equation in the appropriate regimes.

\end{document}